\documentclass[10.5pt,a4paper]{article}
\usepackage{amsmath,amssymb}
\usepackage[pdftex]{graphicx}
\usepackage{latexsym}
\usepackage{slashed}
\usepackage{cite}
\usepackage[hang,bf,figurename=Fig.,font=small]{caption}

\def\Name{}
\def\etal{et~al.}
\newcommand{\Review}[1]{{\textit{#1}}}
\newcommand{\Vol}[1]{{\bfseries{#1}}}
\newcommand{\Year}[1]{(#1)}
\def\Page{}
\def\and{and}
\newcommand{\REVIEW}[4]{{\textit{#1}\ }{#2\ }{#3\ }{#4\ }}
\newcommand{\Book}[1]{{\textit{#1}}}
\def\Publ{}
\usepackage[T1]{fontenc}
\usepackage{tgtermes}
\usepackage{tgheros}
\usepackage{inconsolata}
\usepackage[top=24truemm,bottom=35truemm,left=15truemm,right=15truemm]{geometry}
\title{\bfseries An alternative attractor in gauged NJL inflation}
\author{Tomohiro Inagaki${}^{1,2}$, Sergei D. Odintsov${}^{3,4,5,6}$ and Hiroki Sakamoto${}^7$\\ \\
\small ${}^1$Information Media Center, Hiroshima University, Higashi-Hiroshima, 739-8521, Japan,\\
\small ${}^2$Core of Research for the Energetic Universe, Hiroshima University, Higashi-Hiroshima, 739-8526, Japan,\\
\small ${}^3$Instituci\'{o} Catalana de Recerca i Estudis Avan\c{c}ats (ICREA), Barcelona, Spain,\\
\small ${}^4$Inst. of Space Sciences (IEEC-CSIC), C. Can Magrans s/n, 08193 Barcelona, Spain,\\
\small ${}^5$Tomsk State Pedagogical University, 634061, Tomsk, Russia\\
\small ${}^6$National Research Tomsk State University, 634050, Tomsk, Russia\\
\small ${}^7$Department of Physics, Hiroshima University, Higashi-Hiroshima, 739-8526, Japan}
\begin{document}
\maketitle
\abstract{We have investigated the attractor structure for the CMB fluctuations in composite inflation scenario within the gauged Nambu-Jona-Lasinio (NJL) model. Such composite inflation represents an attractor  which can not be found in a fundamental scalar model. As is known, the number of  inflationary models contains the attractor  classified by the $\alpha$-attractor model. It is found that the attractor inflation in the gauged NJL model corresponds to the $\alpha = 2$ case.}
\normalsize
\section{Introduction}
The early-time inflationary expansion  can successfully solve number of cosmological problems. Some evidence of the inflation can be found in the thermal fluctuations of the Cosmic Microwave Background (CMB). The fluctuations have been precisely observed by the Planck Satellite and parameterized by the curvature perturbation, $A_s$, the spectral index $n_s$, the tensor-to-scalar ratio, $r$ and the running of the spectral index, $\alpha_s$. Many works have been done to construct the particle physics models which satisfy the constraints of the observed fluctuations \cite{Ade:2015xua, Ade:2015tva}. One of models to predict consistent results of CMB fluctuations is the gauged NJL model \cite{Inagaki:2015eza}. It should be noted that there is a possibility to construct a consistent model in a modified gravity, for example, $R^2$ inflation \cite{Starobinsky:1980te} or even its unification with the dark epoch within the modified gravity\cite{nojiri}.

It has been found that the estimated CMB fluctuations are attracted to a stable fixed point at a suitable limit in the broad class of the slow-roll inflation models, including the $R^2$ inflation. The various classes of the inflationary models are classified by the attractor. There are two types of the well-known models, the $\xi$-attractor model  with the non-minimal curvature coupling $\xi$ \cite{Bezrukov:2007ep, Kallosh:2013tua} and the $\alpha$-attractor model which is motivated by supergravity and superconformal theory \cite{Ferrara:2013rsa}. The Higgs inflation belongs to the class of the $\xi$-attractor model where the non-minimal curvature coupling $\xi$ parameterizes the model. The $\alpha$-attractor model is characterized by the parameter $\alpha$ in the K\"{a}hler potential. The potential of the $\xi$-attractor model can be parametrized by $\alpha \equiv 1 + 1/(6\xi)$ due to the special choice of the kinetic term in the Jordan frame\cite{Galante:2014ifa}. In the Einstein frame it is found to be
\begin{align}
  V = V_0 (1 - e^{-\sqrt{2/(3\alpha)}\phi})^{2m},
  \label{eq:e-model}
\end{align}
where $\phi$ is an inflaton field and $\alpha$, $m$ denote parameters of the model. The potential of the $\alpha$-attractor model is given by
\begin{align}
  V = V_0 \tanh^{2m}(\phi/\sqrt{6\alpha}).
  \label{eq:t-model}
\end{align}
Both classes of models are called $\alpha$-attractor models.
In the $\alpha$-attractor models the CMB fluctuations are obtained at the leading order of the $1/N$ expantion,
\begin{align}
  n_s -1 \sim - \frac{2}{N}, \quad r \sim \frac{12\alpha}{N^2}, \quad \alpha_s \sim -\frac{2}{N^2},
  \label{eq:alpha_att}
\end{align}
for $0 < \alpha < O(1)$, where $N$ denotes the e-folding number, which requires $N\sim 50-60$ to solve the horizon and the flatness problems.
Various $\alpha$-attractor models are also investigated (see Ref.~\cite{Kallosh:2015zsa}
for $\alpha=1/3$, Ref.~\cite{Linde:2014hfa, Kallosh:2015lwa} for $\alpha=1/9$ and Ref.~\cite{Odintsov:2016jwr} for the inverse symmetry attractor models). Since a large $\xi$ which may not always being consistent with GUTs predictions for $\xi$ \cite{odintsov} is necessary to obtain a consistent tensor-to-scalar ratio in the Higgs inflation, the model corresponds to the $\alpha$-attractor model with $\alpha=1$ at the large $\xi$ limit. For $\alpha=1$ equal values of the CMB fluctuations are predicted with the $R^2$ inflation \cite{Bezrukov:2007ep}. Ordinary  $\phi^{2m}$ chaotic inflation is also obtained at the large-$\alpha$ limit of the $\alpha$-attractor model\cite{Kallosh:2015zsa, Mosk:2014cba}.

Composite models of the particle physics have been also investigated as a source of the energy density at the early Universe (see, for review \cite{Channuie:2014ysa}, for NJL model \cite{Inagaki:2015eza, Channuie:2016iyy}). The composite model is realized by the condensation of a pair of particles, in which some interactions become very strong like a hadron phase in QCD. A simple model to induce the fermion pair condensation is proposed by Nambu and Jona-Lasinio \cite{Nambu:1961}, and one describes scalar and pseudo-scalar type light mesons as a low-energy effective theory of QCD. The gauged NJL model, the NJL model with QED gauge interaction, has been introduced to deal with QCD low-energy phenonema with QED gauge interaction \cite{Miransky:1993, Harada:2003jx, Kondo:1991yk, Kondo:1992sq, Leung:1985sn, Harada:1994wy}.
The model has been extended to the curved space-time in Refs.~\cite{Geyer:1996kg}.
In our previous papers \cite{Inagaki:2015eza} the gauged NJL model is proposed as an alternative scenario of Higgs inflation and we call the model the gauged NJL inflation.

The main purpose of this letter is to find the attractor behavior of the CMB fluctuations in the gauged NJL inflation and classify it in terms of the $\alpha$-attractor model. This paper is organized as follows. In Sect.~2 we briefly discuss  the gauged NJL model as applied to the inflation. This model is described by the gauge-Higgs-Yukawa theory with a matching condition at the compositeness scale. We apply the slow-roll scenario of the inflation and evaluate CMB fluctuations. In Sect.~3 we derive the attractor of the CMB fluctuations analytically and confirm the attractor behavior by  the numerical calculation. Concluding remarks are given in Sect.~4.
\section{Gauged NJL inflation}\label{sec:2}
The gauged NJL model is considered as an effective theory of a strongly coupled QCD like interaction with an additional $SU(N_c)$ gauge interaction. We assume that the QCD like interaction is strong enough to describe it by four-fermion interactions at inflation era. Then we start from the Lagrangian with $N_f$ flavors of massless fermions,
\begin{align}
        {\cal L}_{gNJL} =& {\cal L}_{gauge} + \bar{\psi} i\hat{\slashed{D}}\psi + \frac{16\pi^2 g_4}{8 N_f N_c \Lambda^2} \left[\left(\bar{\psi}\psi\right)^2+\left(\bar{\psi}i\gamma_5 \tau^{a}\psi\right)^2\right],
        \label{L:gNJL}
\end{align}
where $\mathcal{L}_{gauge}$ represents the pure gauge sector of $SU(N_c)$ gauge symmetry, $\hat{D}_\mu$ denotes the covariant derivative and $\tau^a$ are the generators of the flavor symmetry. The four-fermion coupling is described by the dimensionless parameter, $g_4$ with the number of fermion flavors, $N_f$, the degree of the gauge group, $N_c$, and the compositeness scale, $\Lambda$. Below we employ the $1/N_c$ expansion and take only the leading order terms for simplicity. Note that the perturbative region is given by $\alpha_g \equiv g^2 N_c/(4\pi) \ll 1$ with the $SU(N_c)$ gauge coupling, $g$.

The Lagrangian \eqref{L:gNJL} rewritten by the auxiliary field method can be matched with the renormalization group (RG) improved gauge-Higgs-Yukawa theory at the compositeness condition \cite{Bardeen:1989ds, Hill:1991jc}. Hence, the RG improved effective action of the gauge-Higgs-Yukawa theory is considered as the low energy effective model of the gauged NJL model below the compositeness scale.
Here we assume that the composite scalar field, $\sigma \sim -4\pi^2 g_4 \bar{\psi}{\psi}/(N_fN_c\Lambda^2)$, dominates the energy density of the universe at the inflation era and neglect the other fields. Then the RG improved effective action in curved space-time is given by \cite{Inagaki:2015eza} \begin{align}
        S_J = \int d^4x \sqrt{-g} \left[-\frac{\Omega^2}{2}R + \frac{1}{2}g^{\mu\nu}\partial_\mu \sigma \partial_\nu \sigma - U \right],
\end{align}
with a function, $\Omega^2$, given by
\begin{align}
        \Omega^2 = 1 + \zeta_1 \mu^{2-2n} \sigma^{2n} - \zeta_2\sigma^2,
  \label{eq:Weyl}
\end{align}
and the potential $U$,
\begin{align}
        U = \lambda_1 \mu^2 \sigma^2 + \lambda_2 \mu^{4-4n} \sigma^{4n} - \lambda_3 \sigma^4,
  \label{eq:PtJordan}
\end{align}
where $\mu$ denotes the renormalization scale and we define
\begin{align}
  n \equiv \frac{1}{1 + 3\alpha_g (N_c^2 - 1)/(4\pi N_c)} < 1.
  \label{eq:n}
\end{align}
From the compositeness condition the coefficients $\lambda_1$, $\lambda_2$, $\lambda_3$, $\zeta_1$ and $\zeta_2$ are given by
\begin{align}
        &\lambda_1 = \frac{1-n}{n G_{4r}}\left[\frac{1}{1 - (\mu^2/\Lambda^2)^{\frac{1-n}{n}}}\right], \label{coe:lambda1}\\
        &\lambda_2 = \frac{3-2n}{4n}\left(\frac{N_fN_c}{8\pi^2}\frac{n}{1-n}\right)^{1-2n}\left[\frac{1}{1 - (\mu^2/\Lambda^2)^{\frac{1-n}{n}}}\right]^{2n}, \label{coe:lambda2}\\
        &\lambda_3 = \frac{2\pi^2}{N_fN_c}\frac{1-n}{n} \left[\frac{(\mu^2/\Lambda^2)^{\frac{1-n}{n}}}{1-(\mu^2/\Lambda^2)^{\frac{1-n}{n}}}\right]^2, \label{coe:lambda3} \\
        & \zeta_1 = \frac{1}{6n}\left(\frac{N_fN_c}{8\pi^2}\frac{n}{1-n}\right)^{1-n}\left[\frac{1}{1 - (\mu^2/\Lambda^2)^{\frac{1-n}{n}}}\right]^n, \label{coe:zeta1}\\
        &\zeta_2 = \frac{1}{6}\left[\frac{({\mu}^2/{\Lambda}^2)^{\frac{1-n}{n}}}{1-({\mu}^2/{\Lambda}^2)^{\frac{1-n}{n}}}\right], \label{coe:zeta2}
\end{align}
where $G_{4r}$ is the renormalized four-fermion coupling
\begin{align}
\frac{1}{G_{4r}} \equiv \left[\frac{1}{g_{4}(\Lambda)} - \frac{n}{2n-1}\right] \left(\frac{\mu^2}{\Lambda^2}\right)^{\frac{1-2n}{n}}.
\end{align}
In order to avoid an instability of the potential we consider that the compositeness scale is large enough and eliminate the last term in \eqref{eq:Weyl} and \eqref{eq:PtJordan} during the inflation
\cite{Inagaki:2015eza}. 

Since it is more convenient to calculate the CMB fluctuations in the Einstein frame, we perform the Weyl transformation, $g_{\mu\nu} \to \tilde{g}_{\mu\nu} = \Omega^2(x) g_{\mu\nu}$, and rewrite the action as
\begin{align}
        S_E = \int d^4x \sqrt{-\tilde{g}} \left[-\frac{1}{2}\tilde{R} + \frac{1}{2}\tilde{g}^{\mu\nu}\partial_\mu \phi \partial_\nu \phi - \frac{U}{\Omega^4}\right],
        \label{act:ein}
\end{align}
where we redefine the scalar field $\sigma$ to satisfy
\begin{align}
        \frac{d\phi}{d\sigma} = \sqrt{\frac{1}{\Omega^2} + \frac{3}{2}\frac{1}{\Omega^4}\left(\frac{d\Omega^2}{d\sigma}\right)^2}.
        \label{eq:redef}
\end{align}
in order to keep the canonical form of the kinetic term. Below we evaluate the CMB fluctuations starting from the potential $V\equiv U/\Omega^4$ in the Einstein frame with the relationship (\ref{eq:redef}).

In the slow-roll inflation scenario CMB fluctuations are evaluated by the slow-roll parameters,
\begin{align}
  \varepsilon \equiv \frac{1}{2}\left(\frac{1}{V}\frac{\partial V}{\partial \phi}\right)^2, \quad
  \eta \equiv \frac{1}{V}\frac{\partial^2 V}{\partial \phi^2}, \quad
  \xi \equiv \frac{1}{V^2}\frac{\partial V}{\partial \phi}\frac{\partial^3 V}{\partial \phi^3}.
\end{align}
The field configurations $\phi_{end}, \sigma_{end}$ at the end of the
inflation are determined when one of the slow-roll parameters is equal
to unity.
As it was discussed in \cite{Liddle:1994dx}, the slow-roll era can end at
higher order in terms of the slow-roll expansion, and thus the exit
from inflation can occur. It should be noted that the slow-roll parameter
$\varepsilon$ firstly reaches order unity in the present model.
The field configurations $\phi_N, \sigma_N$ at the horizon crossing are found to generate an enough e-folding number, $\displaystyle N \equiv \int^{\phi_N}_{\phi_{\mathrm{end}}} \frac{V}{\partial V/ \partial\phi} d\phi \sim 50-60$. Then we calculate the spectral index, $n_s$, the running of the spectral index, $\alpha_s$, for the scalar type fluctuation and the tensor-to-scalar ratio, $r$, by
\begin{align}
        & n_s -1 = 2\eta|_{\sigma=\sigma_N} - 6\varepsilon|_{\sigma=\sigma_N}, \label{eq:ns}\\
        & \alpha_s = -24\varepsilon^2|_{\sigma=\sigma_N} + 16\varepsilon\eta|_{\sigma=\sigma_N} - 2\xi|_{\sigma=\sigma_N}, \label{eq:alphas}\\
        & r = 16\varepsilon|_{\sigma=\sigma_N}. \label{eq:r}
\end{align}
We numerically calculate these parameters and discuss the explicit expressions for the attractor.
\section{Attractor solutions in CMB fluctuations}
The CMB fluctuations of the gauged NJL model have been analytically and numerically evaluated in Ref.~\cite{Inagaki:2015eza}. 
The explicit expressions for the spectral index, $n_s$, the running of the spectral index, $\alpha_s$, and the tensor-to-scalar ratio, $r$ have been found for a large e-folding number $N \gg 1$ at a few limits. It has been concluded that the CMB fluctuations coincide with the result in the $\phi^2$ chaotic inflation
\begin{align}
  &n_s -1 = -\frac{4}{2N+1}\sim - \frac{2}{N}, \\
  &r =\frac{16}{2N+1}\sim\frac{8}{N}-\frac{4}{N^2}, \\
  &\alpha_s =-\frac{8}{(2N+1)^2}\sim -\frac{2}{N^2} ,
 \end{align}
at the flat limit $3(d\Omega^2/d\sigma)^2/(2\Omega^2)\ll 1$ for $2n<1$ and $\lambda_1\neq 0$. The result of the $R^2$ inflation is reproduced at the steep limit $3(d\Omega^2/d\sigma)^2/(2\Omega^2)\gg 1$ for $\lambda_1 =0$ and approximately estimated as
\begin{align}
  &n_s -1 \sim - \frac{2}{N}, \quad r \sim \frac{12}{N^2}, \quad \alpha_s \sim -\frac{2}{N^2}
  \label{eq:Cmb_st}
\end{align}
This steep limit solution corresponds to the $\alpha$-attractor model with $\alpha=1$.

Here we derive the explicit expressions for the CMB fluctuations at the weak $\alpha_g$ limit for $n=1$. Taking the limit, $\alpha_g \ll 1$, the potential term in the action (\ref{eq:PtJordan}) and the Weyl factor (\ref{eq:Weyl}) reduce to
\begin{align}
  U &= \frac{3\alpha_g (N_c^2-1)}{4\pi N_c} \left[ \frac{\mu^2}{G_{4r}}\sigma^2 + \frac{4\pi^2}{a}\sigma^4 \right] + O(\alpha_g^2), \label{eq:wk_cpl_U}\\
  \Omega^2 &= 1
  + \frac{1}{6}\left[1 + \frac{3\alpha_g(N_c^2-1)}{4\pi N_c} \left(1 - \ln\left(\frac{a N_c\mu^2}{12\pi\alpha_g(N_c^2-1)\sigma^2}\right)\right)\right]\sigma^2
  + O(\alpha_g^2). \label{eq:wk_cpl_Omg}
\end{align}
It should be noticed that the mass term is negligible and $3(d\Omega^2/d\sigma)^2/(2\Omega^2)\sim 1$ for a large field configuration $\sigma\gg 1$. The potential term of the action  (\ref{act:ein}) is simplified to
\begin{align}
  V=\frac{U}{\Omega^4} \rightarrow \frac{108\alpha_g (N_c^2-1)}{aN_c}(1-\exp(-\sqrt{1/3}\phi))^2,
  \label{eq:wpot}
\end{align}
for $\sigma \gg 1$. Since a small $\sigma$ part of the potential is essential to determine the end of the slow roll, the e-folding number $N$ cannot be directly estimated from \eqref{eq:wpot}.  The large field  $\Delta \phi \equiv |\phi_N - \phi_{\mathrm{end}}| \sim O(10)$ in the gauged NJL inflation allows to neglect the contribution from the end of the slow roll described by $\phi_{\mathrm{end}}$. In such an approximation the field configuration at the horizon crossing is  given by
\begin{align}
 \phi_N \sim \sqrt{3}\ln \frac{2N}{3} .
\end{align}
Substituting this configuration into Eqs.~(\ref{eq:ns}), (\ref{eq:alphas}) and (\ref{eq:r}) we obtain
\begin{align}
  n_s -1 \sim - \frac{2}{N}, \quad r \sim \frac{24}{N^2}, \quad \alpha_s \sim -\frac{2}{N^2},
  \label{eq:Cmb_wk_cpl}
\end{align}
up to the leading order of the $1/N$ expansion. Recalling the predictions of $\alpha$-attractor model \eqref{eq:alpha_att}, the results at the weak-coupling limit correspond to the $\alpha$-attractor model with $\alpha=2$. This means the gauged NJL inflation belongs to the special class of the universal attractor model and is distinguishable from the Higgs type scalar inflation.

To clarify the attractor behavior of the CMB fluctuations we fix the model parameters, $G_{4r}$, $\alpha_g$, $N_c$, $N_f$, $\Lambda$, $\mu$ and numerically calculate the spectral index, $n_s$, the running of the spectral index, $\alpha_s$, and the tensor-to-scalar ratio, $r$.  After the systematic study of each model parameter dependence of the CMB fluctuations, we observe that the solution approaches the steep limit as the number of the fermion flavors $N_f$ increases. The solution approaches the weak-coupling limit as the gauge coupling $\alpha_g$ decreases. We numerically show the typical attractor behavior.

\begin{figure}[h]
  \hfill
  \begin{minipage}[b]{0.4\hsize}
    \includegraphics[width=\linewidth]{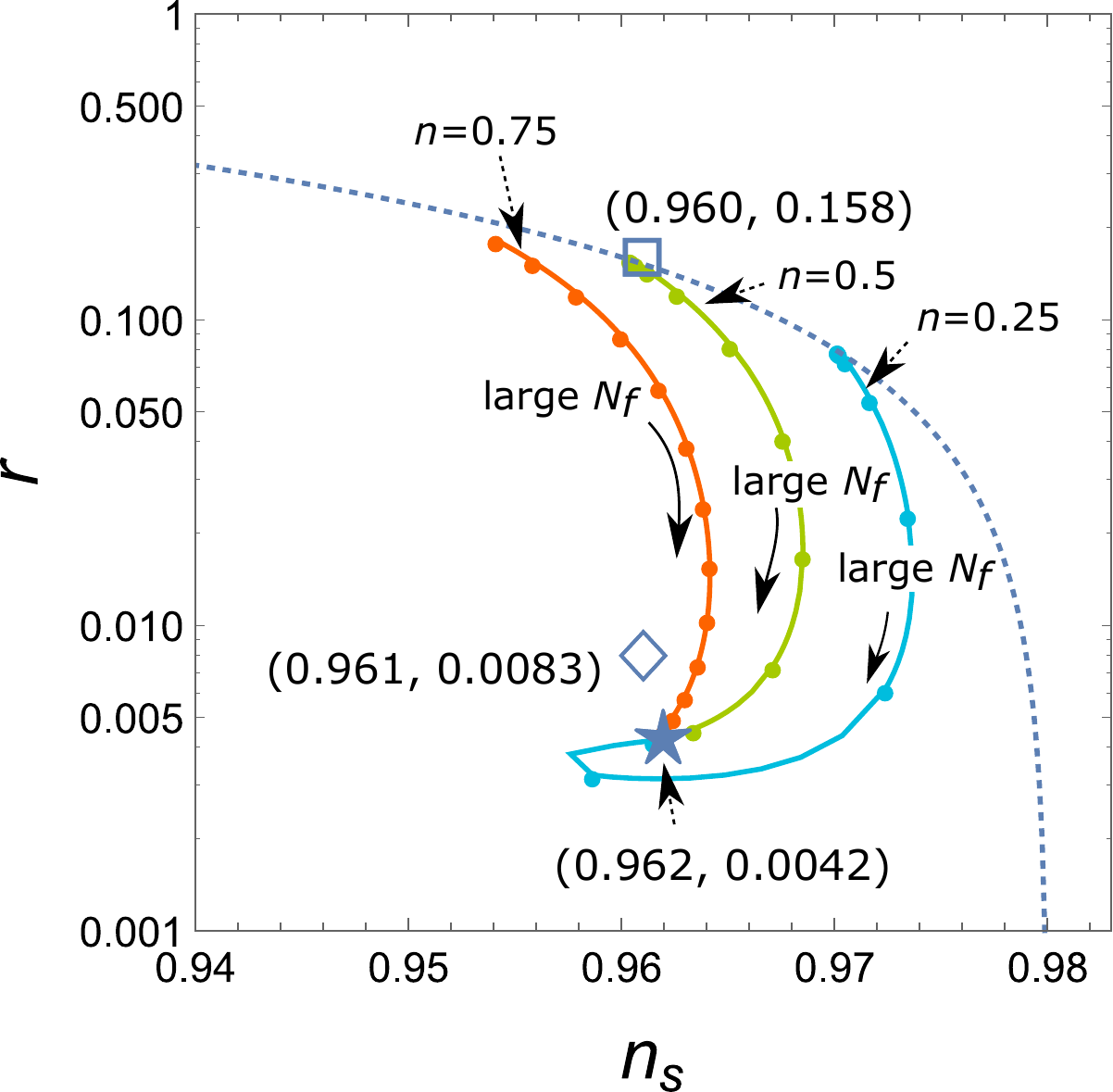}
  \end{minipage}
  \noindent\hfill
  \begin{minipage}[b]{0.43\hsize}
    \includegraphics[width=\linewidth]{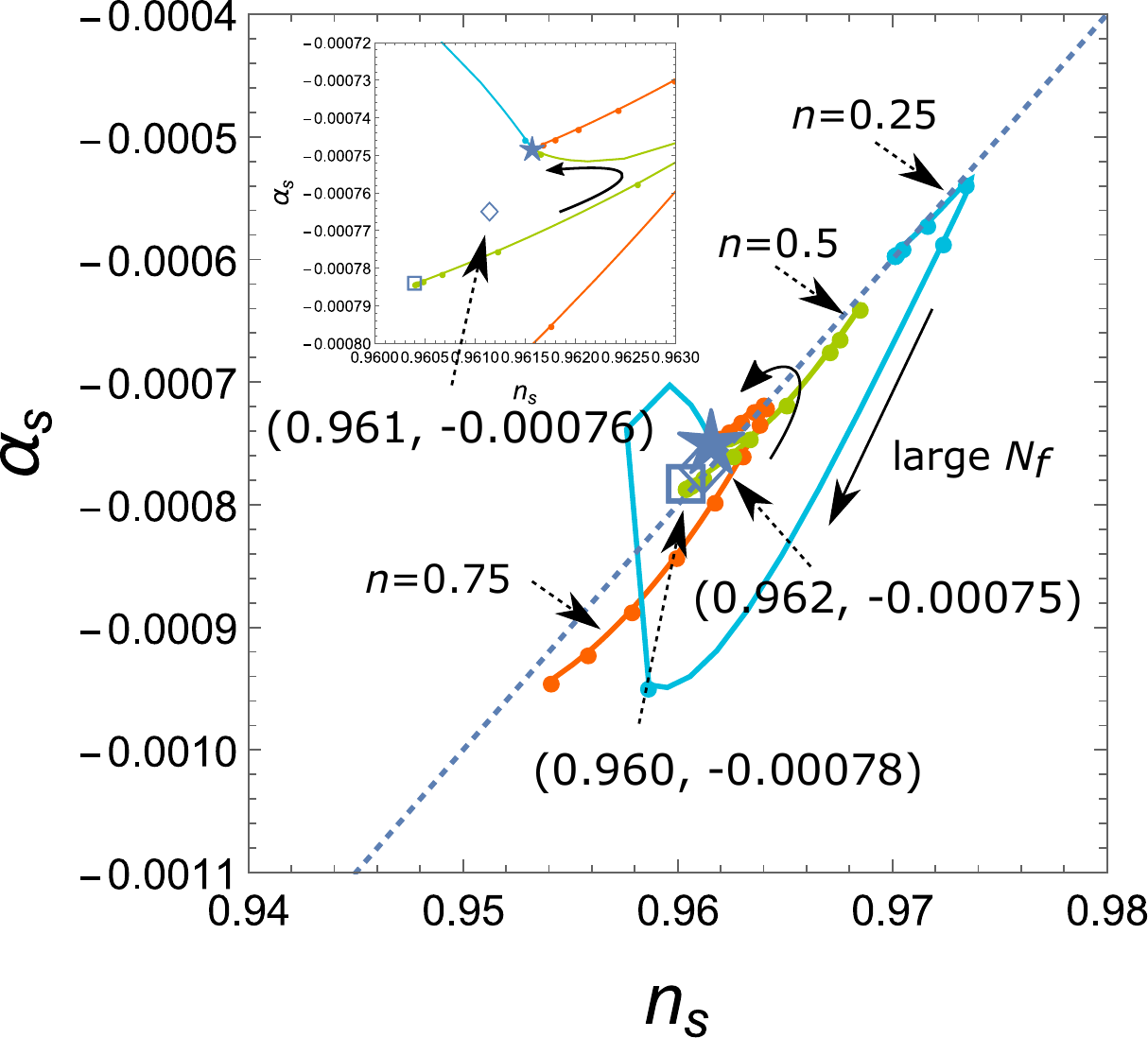}
  \end{minipage}
  \noindent\hfill\null
  \caption{The $N_f$ dependence of $(n_s,r)$ and $(n_s,\alpha_s)$ for $G_{4r} = 10^{10}$, $N_c = 3$ and $\mu = 10^{-3}$. Each point on the solid lines shows the prediction for $N_f = (10^0, 10^1, \cdots, 10^{14})$. The symbols $\Box$, $\star$, and $\diamond$ denote the points in table \ref{tab:tab1}. The dotted lines show the prediction by the $\phi^{4n}$ chaotic inflation with varing $n$.}
  \label{fig:u-att}
\end{figure}

In Fig.~\ref{fig:u-att} we set $G_{4r} = 10^{10}$, $N_c=3$, $\mu =
10^{-3}$ and plot the $N_f$ dependence of $(n_s,r)$ and
$(n_s,\alpha_s)$ with varying the number of fermion flavors $N_f$.
Here we use the parameter, $n$. The gauge couploing $\alpha_g$ is
ditermined by Eq.~\eqref{eq:n}.
The trajectories illustrated by the solid lines represent the $N_f$ dependence for $n=(0.75, 0.5, 0.25)$. The trajectories start near the dotted line, the prediction by the $\phi^{4n}$ chaotic inflation. It is observed that all trajectories approach the point denoted by the symbol $\star$ at the large-$N_f$ limit. It should be noticed that the effective non-minimal curvature coupling $\zeta_1$ in Eq.~(\ref{coe:zeta1}) becomes large, if we take a large number of flavor $N_f$. A similar behavior is observed in the universal attractor models \cite{Kallosh:2013tua}.  In the Higgs inflation model the solution approaches the same point at the strong scalar-curvature coupling limit.

\begin{figure}[h]
  \centering
  \hfill
  \begin{minipage}[b]{0.4\hsize}
    \centering
    \includegraphics[width=\linewidth]{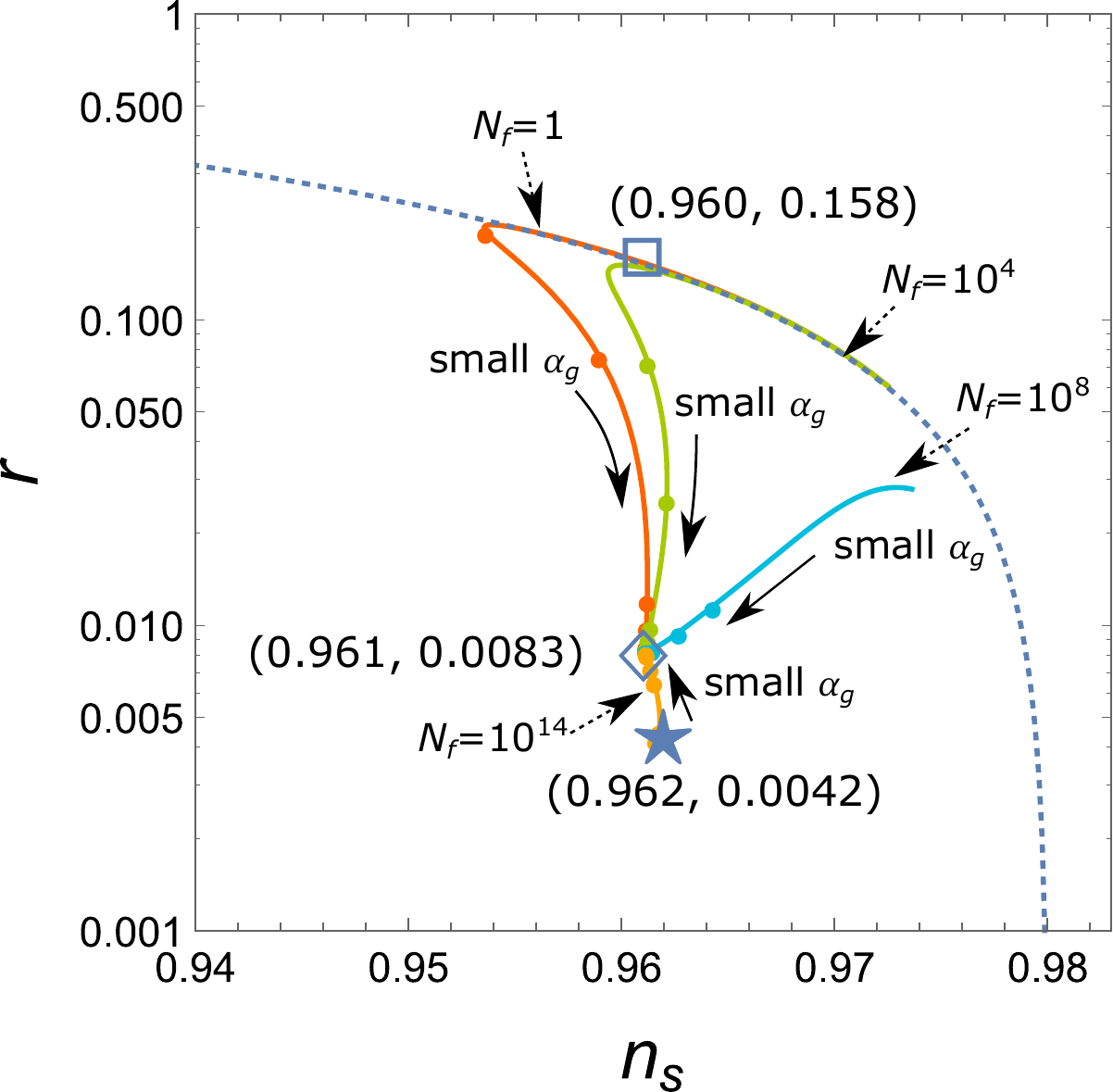}
  \end{minipage}
  \noindent\hfill
  \begin{minipage}[b]{0.43\hsize}
    \centering
    \includegraphics[width=\linewidth]{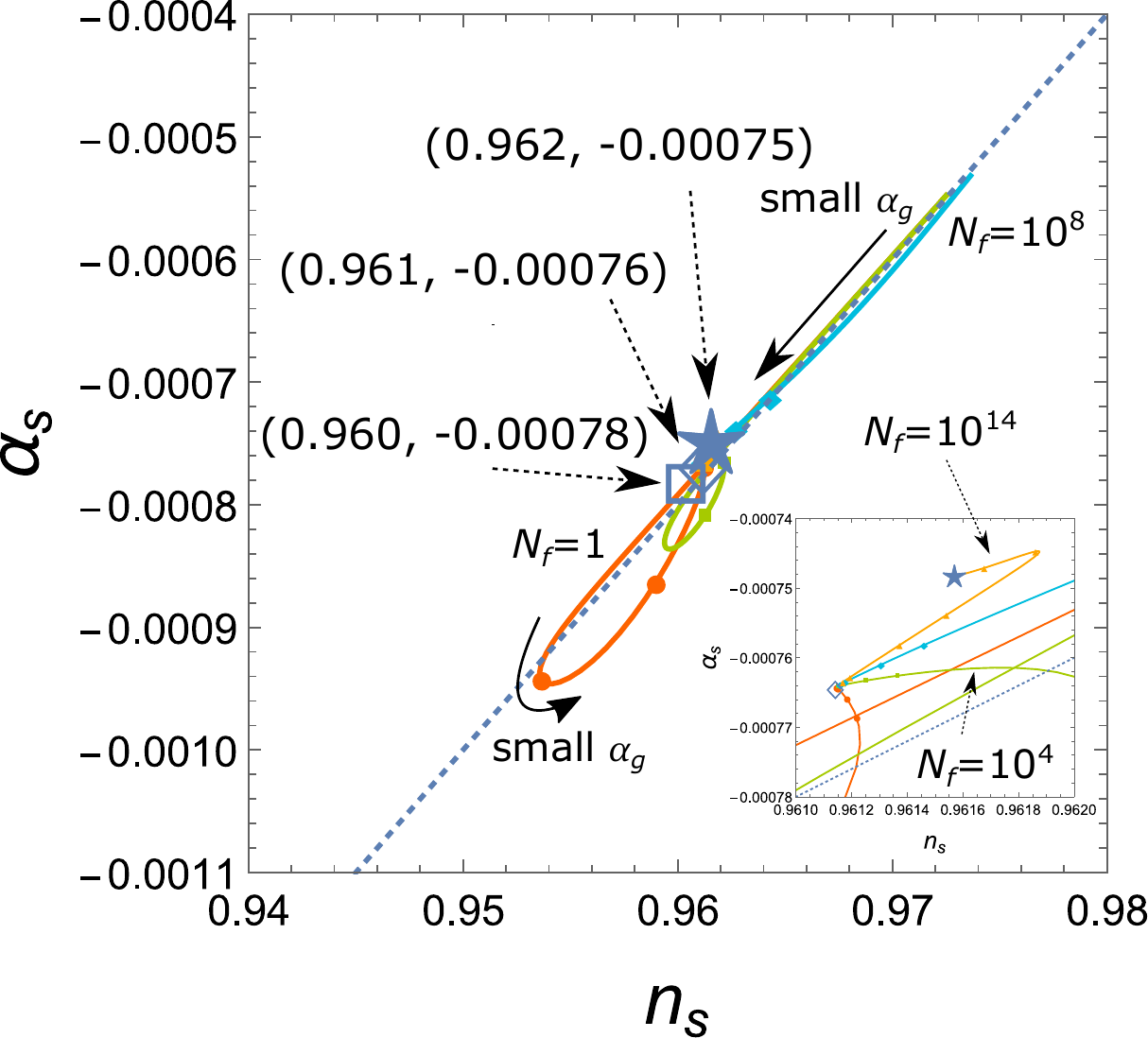}
  \end{minipage}
  \hfill\null
  \caption{The $\alpha_g$ dependence of $(n_s,r)$ and $(n_s,\alpha_s)$ for $G_{4r} = 10^{10}$, $N_c=2$ and $\mu = 10^{-3}$. Each point on the solid lines shows the prediction for $\alpha_g = (1, 0.5, 0.1, 0.05, 0.01, 0.005)$. The symbols $\Box$, $\star$, and $\diamond$ denote the points in table \ref{tab:tab1}. The dotted lines show the prediction by the $\phi^{4n}$ chaotic inflation with varing $n$.}
  \label{fig:g-att}
\end{figure}

In Fig.~\ref{fig:g-att} we illustrate the gauge coupling dependence of $(n_s,r)$ and $(n_s,\alpha_s)$ for $G_{4r} = 10^{10}$, $N_c=2$ and $\mu = 10^{-3}$. The solid lines show the attractor behavior in the gauged NJL model for $N_f = (1, 10^4, 10^8, 10^{14})$. For a strong coupling $\alpha_g\sim 1$ the trajectories flow along the dotted line, the prediction by the $\phi^{4n}$ chaotic inflation. The tendency is remarkable for a small $N_f$. Then all the trajectories leave for the symbol $\diamond$ as $\alpha_g$ decreases. First we set a large $N_f$ and then take the small $\alpha_g$ limit, we observe the flow from $\star$ to $\diamond$. It should be noticed that there is no opposite direction flow from $\diamond$ to $\star$. The attractor indicated by the symbol $\diamond$ is characteristic for the gauged NJL inflation which is distinguishable from the universal attractor model.

In table ~\ref{tab:tab1} we show the spectral index, $n_s$, the running of the spectral index, $\alpha_s$, and the tensor-to-scalar ratio, $r$, at the flat, the steep and the weak-coupling limit for $N=50$. The flat limit solutions are well evaluated in the leading order of the large-$N$ limit. A finite contribution from $\phi_{end}$ makes some discrepancy between the numerical and large-$N$ results for the steep and the weak-coupling limit. The results at the steep and the weak-coupling limit correspond to the $\alpha$-attractor model with $\alpha=1$ and $2$, respectively, and satisfy the observational constraints \cite{Ade:2015xua, Ade:2015tva}.

\begin{table}[h]
  \caption{Fixed point for the CMB fluctuations for $N=50$ in the gauged NJL model}
  \label{tab:tab1}
  \centering
  \begin{tabular}{lll|lll}
    & & & $n_s$ & $r$ & $\alpha_s$ \\
  \hline
    & & Flat limit ($2n<1$) & 0.960 & 0.158 & -0.00080 \\
    Large $N$  & & Steep limit  & 0.960 & 0.0048 & -0.00080 \\
    & & Weak coupling limit  & 0.960 & 0.0096 & -0.00080 \\
  \hline
    & $\Box$ & Flat limit ($2n<1$) & 0.960 & 0.158 & -0.00078 \\
    Numerical  & $\star$ & Steep limit &0.962 & 0.0042 & -0.00075 \\
    & $\diamond$ & Weak coupling limit & 0.961& 0.0083 & -0.00076
  \end{tabular}
\end{table}
\section{Conclusion}
We have studied the attractor behavior of $n_s$, $r$, and $\alpha_s$ in the gauged NJL inflation. Applying the auxiliary field method and improving the action by the RG equation at the leading order of the $1/N_c$ expansion, we obtain effective potential \eqref{eq:PtJordan} with the Weyl factor  \eqref{eq:Weyl}. The effective action contains the mass, $\phi^{4n}$ interaction, and the non-minimal curvature coupling terms. We adopt the slow roll scenario of the chaotic inflation and calculate  the spectral index, $n_s$, the running of the spectral index, $\alpha_s$, and the tensor-to-scalar ratio, $r$.

The effective curvature coupling $\zeta_1$ can play the role of the attractor parameter $\xi$ in the $\xi$-attractor model. If we take a large number of flavors $N_f$, the effective curvature coupling $\zeta_1$ becomes large. Hence the predictions of CMB fluctuations approach the universal attractor at $\alpha=1$. If we take the weak-coupling limit, $\alpha_g\rightarrow 0$, we observe an alternative attractor behavior corresponding to $\alpha=2$ in the $\alpha$-attractor model. The flows of the solution are presented in Figs.~\ref{fig:u-att} and \ref{fig:g-att}. We observe a flow from the attractor at $\alpha=1$ to the one at $\alpha=2$ and the flow with the opposite direction is not possible.

We often take the attractor parameter $\alpha$ to be unity because the large-curvature coupling requires to satisfy the observed constraint for the power spectrum of the curvature perturbation. However, in the gauged NJL inflation, the effective curvature coupling, $\zeta_1 \sim 1/6$, satisfies the constraint. This means that we can construct the relevant model corresponding to the $\alpha$-attractor model with $\alpha = 2$. Thus the gauged NJL inflation is comparable to the Higgs inflation and has a specific attractor.

Although the present work is restricted to the analysis of the properties of the gauged NJL inflation, it points to the a possibility to  constrain the model parameters by observing physics at the early Universe.
\section*{Acknowledgements}
The work by TI is supported in part by JSPS KAKENHI Grant Number 26400250 and
that by SDO is supported in part by MINECO (Spain), project FIS2016-76363-P and by CSIC I-LINK1019 project and by Russ. Min. of Education and Science, project No. 3.1386.2017.

\end{document}